\documentclass[11pt,twoside]{article}
\usepackage{cozumel2005}
\usepackage{epsf}
\usepackage{graphicx}
\usepackage{lscape}
\begin{document}
\title{On the Calibration of the Temperatures and Colors of Stellar
   Models for Lower Mass Stars}    
\author{Don A.~VandenBerg}   
\affil{Dept.~of Physics \& Astronomy, University of Victoria, Box 3055,
   Victoria, B.C., Canada~~~V8W~3P6}    

\begin{abstract} 
The temperatures and colors of stellar models are much less secure than
predicted luminosities because they depend sensitively on the still very
uncertain physics of stellar envelopes and atmospheres.  Consequently, it
is important to ensure that the models which are used to interpret stellar
populations data satisfy existing observational constraints on these
properties.  As shown in this study, the available $T_{\rm eff}$ contraints
do not pose a problem for evolutionary calculations in which the free parameter
($\alpha_{\rm MLT}$) in the Mixing-Length Theory of convection (which continues
to be widely used), is set to the value required by a Standard Solar Model.
That is, it is still not possible to say whether some dependence of
$\alpha_{\rm MLT}$ on mass, metallicity, or evolutionary state should be
invoked.  On the other hand, the evidence seems compelling that adjustments
of the color transformations from model atmospheres are needed to achieve 
consistency with observations of cool stars.  Stellar models that have been
normalized to the Sun and transformed to the observed plane using recent
empirical color--$T_{\rm eff}$ relations are able to provide superb fits to
the M$\,$67 and Hyades color-magnitude diagrams (CMDs).  Because the properties
of metal-poor stars and star clusters (metallicities, distances, etc.) are
more uncertain, it is much harder to constrain the $T_{\rm eff}$ and color
scales at low metallicities than at [Fe/H] $\approx 0.0$.  These difficulties
are highlighted in the following examination of the constraints provided by
Population II subdwafs and several well-observed globular clusters (M$\,$3,
M$\,$5, M$\,$68, M$\,$92, and 47 Tuc).  It is not clear that significant
improvements to current semi-empirical color--$T_{\rm eff}$ relations
for metal-deficient stars will be possible until much tighter observational
constraints become available.  Brief comparisons of the evolutionary tracks
computed by different workers are also included in this investigation.


\end{abstract}


\section{Introduction}                      
Our understanding of stellar populations is only as good as the models that are
used to interpret the data and the accuracy to which such basic properties as
distance and metallicity are known (assuming that the observations are
reliable).  Due to the many factors at play, and the many degeneracies between
them, it is virtually impossible to find a unique solution to the values of
the various quantities of interest simply by incomparing synthetic and observed
color-magnitude diagrams (CMDs).
 
The lack of a rigorous theory for super-adiabatic convection, in particular,
makes it very risky to place too much reliance on predicted effective
temperatures, in both an absolute and relative sense (especially at
metallicities quite different from solar).  This is why there has long been a
general consensus that ages are best derived from computed turnoff (TO)
luminosity versus age relations (because TO luminosities are predicted to be
nearly independent of convection theory; e.g., see \citealt{van83}).  Moreover,
it has become clear that there must be some process(es) (e.g., turbulence) at
work in the envelopes of lower mass, metal-poor stars as, otherwise, it is not
possible to explain either the variation of Li with $T_{\rm eff}$ in field halo
dwarfs (see \citealt{rmr02}) or the lack of any detectable difference in the
chemical compositions of TO and red-giant-branch (RGB) stars in such globular
clusters (GCs) as NGC$\,$6397 and 47 Tucanae (see \citealt{gra01}).  The extent
to which diffusive processes are inhibited has ramifications for the predicted
difference in $T_{\rm eff}$ between the TO and lower RGB (\citealt{vrm02}) that
are quite similar to those caused by variations in age, [Fe/H], and [O/Fe]
(e.g., \citealt{vb01}).  This makes the interpretation of the length (and slope)
of the subgiant branch very complicated indeed.

Further uncertainties, on the theoretical side, can be attributed to the
low-temperature opacities, which can alter both the location and the slope of
the RGB, and to the treatment of the atmospheric layers.  It is conventional to
adopt either a gray or scaled-solar $T$--$\tau$ relation to describe the
atmospheric structure and to integrate the equation of hydrostatic equilibrium
in conjunction with this relation to derive the pressure boundary condition that
is used to construct stellar models.  Such a procedure is bound to lead to
systematic errors in the predicted effective temperatures because a fixed
$T$--$\tau$ relation is certainly inappropriate for stars of all masses,
metallicities, and evolutionary states.  A better approach would be to use
proper model atmospheres to provide the outer boundary conditions for stellar
models, as well as the color--$T_{\rm eff}$ relations that are used to transpose
the models from the theoretical to the observed plane.  However, this opens the
door to the many uncertainties inherent in model atmosphere computations and
thereby to further sources of systematic error.  It is clear from the plots
provided by \citet{vc03} that the recent color transformations based, in part,
on model atmospheres by \citet{lcb98}; \citet{cas99}; and \citet{hbs00} differ
considerably from each other, but it is not obvious whose calculations are the
most realistic ones.

Added to this depressing situation is the fact that the photometric data
themselves may have both zero-point and systematic errors; and certainly the
reddenings, distances, and chemical compositions of even the simplest stellar
populations (like GCs) are not as well constrained as they need to be.  For
instance, as shown by \citet{dc99}, current predictions for the luminosities
of horizontal-branch (HB) stars, which are one of the favored ``standard
candles" for distance determinations, do not agree at all well.  Moreover, it
is not uncommon to find 0.2--0.4 dex variations in the [Fe/H] (and [O/Fe])
values derived by different workers.  In the case of NGC$\,$4590 (M$\,$68),
for example, recent metallicity estimates range from [Fe/H] $=-1.99$
(\citealt{cg97}) to $-2.43$ (\citealt{ki03}).  Such differences can have huge
implications for the comparisons of isochrones with photometric data.
  
From time to time, suggestions are made that completely objective techniques
should be used to deduce the properties of stellar populations.  Perhaps the
first such suggestion was made by \citet{fj82}, who devised a statistical
``goodness of fit" criterion to determine the values of $m-M$, age, [Fe/H], $Y$,
and $\alpha_{\rm MLT}$ (where the latter is the usual mixing-length parameter)
from fits of isochrones to the main sequence (MS), TO, and subgiant (SGB)
portions of observed CMDs.  Not surprisingly, given the many potential problems
described above, they concluded that, ``in practice, the results are more
sensitive to unknown calibration effects, and to uncertainties in the
theoretical modelling of convection and stellar atmospheres, than to formal
errors in the fits".  One should not expect meaningful results to be obtained
if the sizes and shapes of isochrones play the dominant role in determining
which isochrone provides the best match to a given CMD, whether or not a
least-squares fitting method is used.

Similar ideas have been proposed recently by \citet{wh03}, but with the 
important difference that the information which is subjected to a least-squares
analysis is the variation in the numbers of stars along the principal sequences
of an observed CMD.  This variation reflects the changes in the evolutionary
timescales in different parts of the H--R diagram, which is widely considered 
to be a more robust prediction of stellar models (with good reason) than, in
particular, surface temperatures.  However, we note that theoretical luminosity
functions (LFs) appear to have such a weak sensitivity to [Fe/H], [$\alpha$/Fe],
and age (see \citealt{vld98}) that they cannot be used to determine these
quantities. (As shown in the same study, LFs may provide a way of determining
the helium abundance in a star cluster or of detecting whether member stars
have been able to retain significant rotational angular momenta throughout
their evolution.)

There was some hope that the color distributions of subgiant-branch stars could
be used to determine distance-independent absolute ages for globular clusters
(see \citealt{bv97}), but a follow-up study by P.~A.~Bergbusch (in preparation)
has found that, here too, the sensitivity of so-called ``color functions" (CFs)
to quantities of interest is likely too low to make this approach competitive
with other methods.  For instance, preliminary indications are that GC ages
derived from CFs have $1\sigma$ uncertainties of at least $\sim \pm 3$ Gyr.  As
a result of such work, it seems doubtful that any major breakthroughs will be
forthcoming from efforts (like that by \citealt{wh03}) which are based on
predictions of the numbers of stars along isochrones.

The simplest stellar populations (open clusters, GCs) provide the best tests of
stellar models (along with well observed binary stars, and field stars with
accurately determined distances and chemical compositions), and they will
continue to do so as our understanding of the physics of stars and the basic
properties of such stellar systems improves over time.  Indeed, it is largely
as a result of work on star clusters that the deficiencies of current stellar
models are revealed (e.g., the need for convective core overshooting in models
of intermediate-mass stars; see the review by \citealt{chi99}).  They also
provide the means to ``calibrate" stellar models so that the latter predict,
for instance, the right MS and RGB slopes.  It makes no sense to try to
interpret the observations for complex stellar populations, such as the
Magellanic Clouds and dwarf spheroidal galaxies, using stellar models that are
incapable of reproducing the CMDs of simple stellar populations to within some
reasonable tolerance.

It is the purpose of the present paper to review recent work that has been
carried out at the University of Victoria to try to produce a set of models
(\citealt{vbd05}) having well constrained temperatures and colors (see VC03),
in the hope that they will offer new and/or improved insights into the stellar
content of complex stellar populations.   Hopefully this work will provide an
instructive example of the degree to which predicted temperatures can be
trusted and of how observations may be used to calibrate the colors of stellar
models for lower mass stars (in particular) over a wide range in metal
abundance.

\section{The Calibration of Stellar Models for [Fe/H] $\approx 0.0$}
The Hyades has long been a pillar of stellar astronomy and of efforts to 
establish the extragalactic distance scale because its distance can be
determined by direct geometric means.  As a result of {\it Hipparcos} secular
and trigonometric parallaxes, its color--$M_V$ diagrams have been established
to unprecedented precision over $\sim 6$ magnitudes of its main sequence (see
\citealt{dhd01}; VC03).  Moreover, both its metallicity ([Fe/H] $\approx +0.13$)
and helium abundance (just slightly subsolar) are known to high accuracy; the
former from high-resolution spectroscopy (e.g., \citealt{bf90}) and the latter
from analyses of its binary stars (\citealt{lfj01}; VC03).  A further advantage
of the Hyades over many other systems is that it suffers from negligible
foreground reddening.  Because so many of its properties involve little
uncertainty, the Hyades obviously provides a fundamental test of stellar models.
Its main deficiency in this regard is that it is too young to have an extended
giant branch, and consequently, does not contrain models for low-gravity stars.

For this reason, and because Hyades stars are super-metal-rich, VC03 opted to
rely on M$\,$67 to derive the color--$T_{\rm eff}$ relations for stars having
[Fe/H] $\approx 0.0$, and to use the Hyades as a secondary check (see below) of
the color transformations so obtained.  M$\,$67 has a well-defined CMD
(\citealt{mmj93}), its metallicity is close to [Fe/H] $=-0.04$ according to the
results of high-resolution spectroscopy (\citealt{ht91}; \citealt{tet00}), and 
there is general agreement that its reddening is within 0.01 mag of $E(B-V) =
0.04$ (\citealt{ntc87}; \citealt{sfd98}; \citealt{svk99}).  The cluster helium
content is not known, but presumably it is close to that of the Sun given the
similarity in metal abundance.  The calculation of a Standard Solar Model
provides the default value of $Y$ for solar abundance stars (as well as the
value of the mixing-length parameter, $\alpha_{\rm MLT}$, that is assumed for
models of stars in all other parts of the H-R diagram).  This estimate of
$Y_\odot$, together with an adopted helium enrichment law (e.g.,
$\Delta Y/\Delta Z = 2.2$; see VandenBerg et al.~2005) provide the means to
determine the helium abundance that is to be assumed at other metallicities.

\begin{figure}[!ht]
\plotone{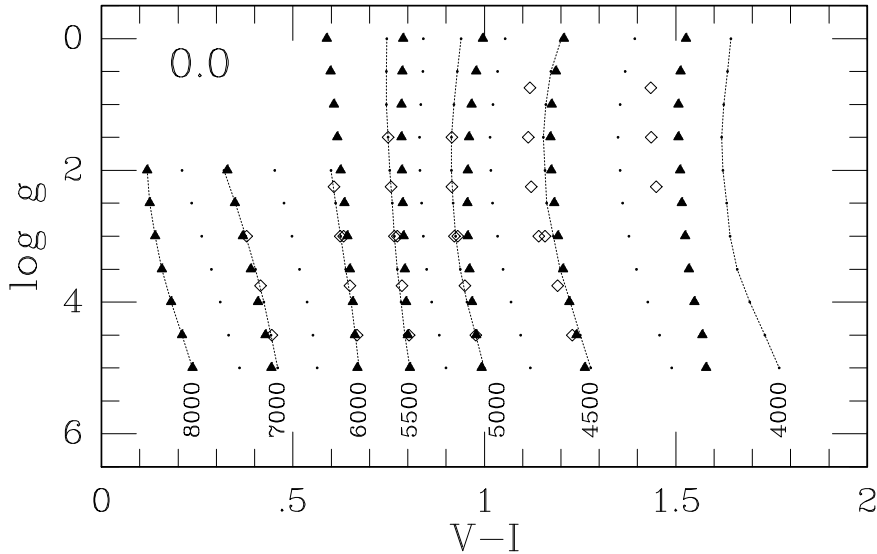}
\caption{Plot of the $(V-I)$--$T_{\rm eff}$--$\log g$ relations that were
adopted by VC03 for stars having [Fe/H] $=0.0$ (small {\it filled circles},
some connected by dotted curves to indicate lines of constant temperature).
The colors predicted by MARCS (VandenBerg \& Bell 1985; Bell \& Gustafsson 1989)
and Kurucz (Castelli 1999) model atmospheres are denoted by the {\it diamond}
symbols and {\it filled triangles}, respectively.}
\end{figure}

As the blanketing in the $B$ bandpass is much more severe than in $V$ and $I$,
one expects that the $B-V$ colors predicted by model atmospheres will be less
trustworthy than $V-I$.  Indeed, it is comforting to find that the latter, when
derived from MARCS (\citealt{vb85}; \citealt{bg89}) and Kurucz (\citealt{cas99})
model atmospheres, are in very good agreement for dwarf stars having [Fe/H]
$=0.0$ and temperatures above $\approx 5000$ K. This is illustrated in
Figure 1.  Furthermore, when the color transformations (from either source) are
applied to models that allow for convective core overshooting, it is found that
a 4.0 Gyr isochrone for [Fe/H] $=-0.04$ (and $Y=0.2735$) provides a very good
match to the upper MS and TO of M$\,$67 (see Figure 2, which is reproduced
from VC03).  In order for that isochrone to provide a good fit of the entire
CMD, the model-atmosphere-based colors for lower MS and RGB stars were
arbitrarily corrected, as necessary.  (However, as discussed below, there is
considerable justification for such adjustments.)

\begin{figure}[!ht]
\plotone{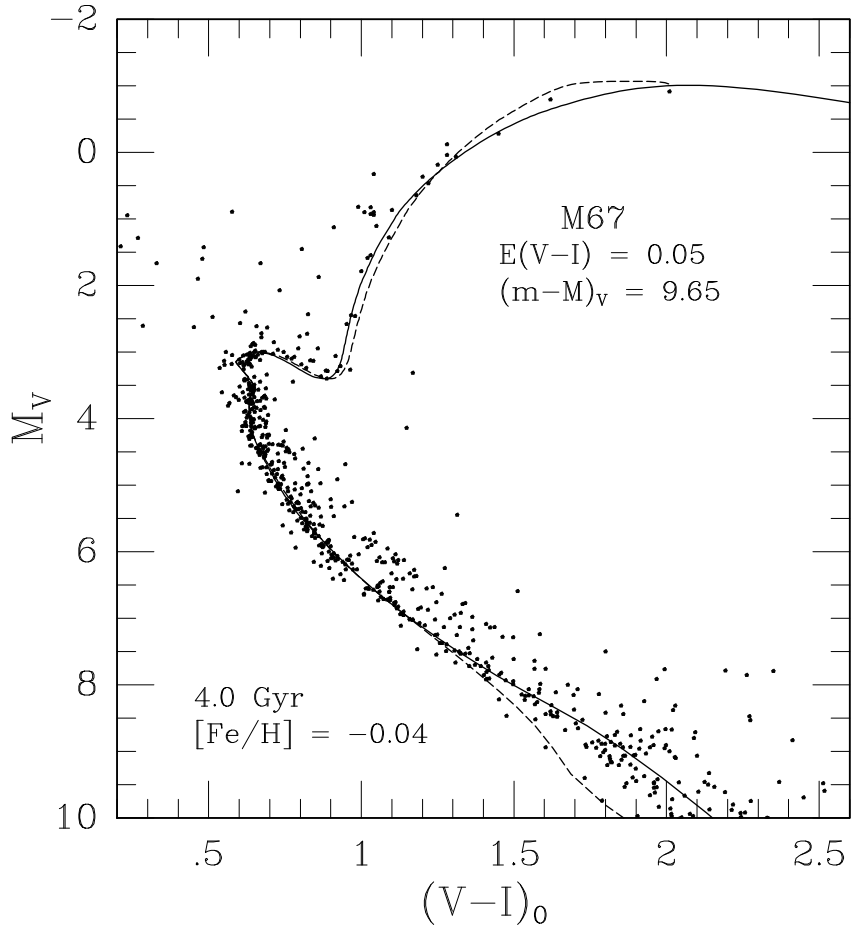}
\caption{Fit of a 4.0 Gyr isochrone for [Fe/H] $=-0.04$ to the $VI$ photometry
of M$\,$67 by Montgomery et al.~(1993), on the assumption of the indicated
reddening and distance modulus.  The {\it solid curve} assumes the color
transformations given by VC03, which are consistent with the predictions from
MARCS model atmospheres (VandenBerg \& Bell 1985; Bell \& Gustafsson 1989)
only at temperatures $\ge 5000$ K (see Fig.~1), whereas the {\it dashed curve}
is obtained if Castelli (1999) color--$T_{\rm eff}$ relations are used.}
\end{figure}  
 
VC03 found that the fit of the same isochrone to $BV$ observations of M$\,$67
(by Montgomery et al.~1993) was much more problematic: the synthetic colors
based on MARCS model atmospheres seemed to give results that were consistent
with the fit to $VI$ data only very near the turnoff, while those derived from
Kurucz model atmospheres were too red.  In order for the 4 Gyr isochrone to
reproduce the observed [$(B-V)_0,\,M_V$]--diagram, VC03 chose to apply whatever
corrections to the theoretical color transformations seemed to be necessary.
However, it turned out that the resultant $(B-V)$--$T_{\rm eff}$ relation for
the MS stars in M$\,$67 agrees very well with what is perhaps the best empirical
relation which has been derived to date for dwarfs having close to the solar
metallicity (that by \citealt{sf00}).  As shown in Figure 3, there are no
significant differences between the two relationships over a large range in 
temperature (or color).  Note that the values of $(B-V)_\odot$ implied by the
VC03 and Sekiguchi \& Fukugita color transformations are, respectively, 0.637
(cf.~\citealt{vp89}) and $0.626\pm 0.018$.  

\begin{figure}[!ht]
\plotone{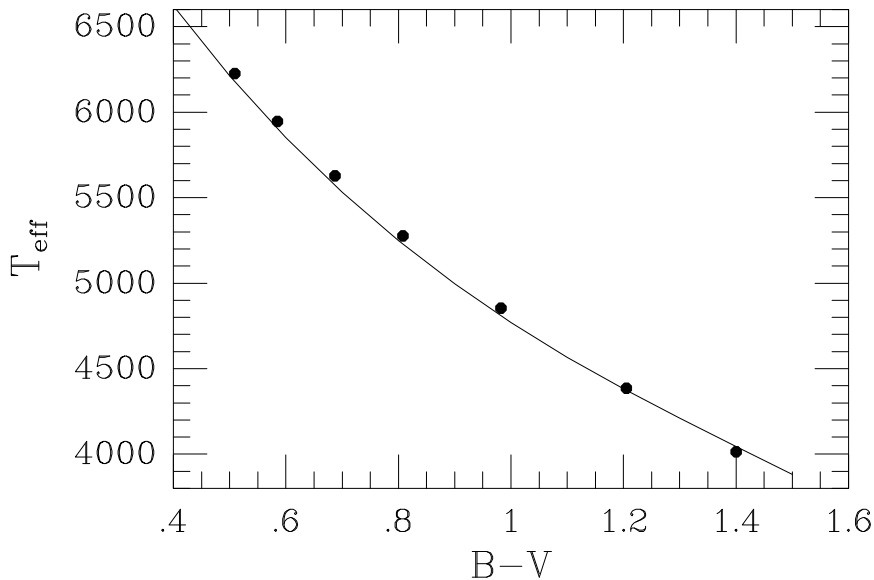}
\caption{Comparison of the $B-V$--$T_{\rm eff}$ relation derived by Sekiguchi
\& Fukugita (2000; {\it solid curve}) for [Fe/H] $=0.0$ with that adopted by
VC03.  The {\it filled circles} represent computed ZAMS models for 0.6--$1.2
{{\cal M}_\odot}$ stars having the solar metallicity.}
\end{figure}

Although the superb fit of the 4 Gyr isochrone to the M$\,$67 CMD obtained by
VC03 (see their Fig.~4) was contrived, a virtually identical fit to the lower
main sequence would be obtained using the \citet{sf00} transformations.  Thus,
the adjustments that VC03 applied to the synthetic colors in order for the
models to match the MS of M$\,$67 are not at all {\it ad hoc}: they are
{\it needed}, in fact, to satisfy empirical constraints.  It then follows that,
in order to reproduce the observed $B-V$ versus $V-I$ relation (i.e., as defined
by the cluster dwarf stars), the corrections to the synthetic $V-I$ colors
adopted by VC03 (essentially the difference between the solid and dashed curves
in Fig.~2) are also necessary.  [Other considerations went into the
determination of the color transformations for the faintest MS stars (see
VC03), which resulted in the isochrone that was fitted to the M$\,$67 CMD being
too red at $M_V \ge 8$ (see Fig.~2).  The correct explanation for this 
discrepancy is not clear.]

As far as the transformations given by VC03 for giants are concerned, it is
evident in Fig.~2 that the difference between the adopted $(V-I)$--$T_{\rm eff}$
relations and those derived from Kurucz model atmospheres are not large (except
near the RGB tip).  Moreover, as shown by VC03 (see their Fig.~27), the
predicted temperatures along the isochrone used to fit the M$\,$67 CMD agree
very well with those derived from the empirical relations between $V-K$ and
$T_{\rm eff}$ by \citet{bcp98} and \citet{vb99}.  Since there are no obvious
problems with the predicted temperatures, the adjustments to the synthetic
$B-V$ and $V-I$ color indices adopted by VC03 in order for the models to match
the colors of the cluster giants are readily justified.  Strong support for
this {\it calibration} of of the colors for solar metallicity stars came from
a follow-up study by \citet{cvg04}, who found that they were able to obtain a
completely consistent fit of the same isochrone to $uvby$ data for M$\,$67
using their semi-empirical transformations to the Str\"omgren photometric
system.  The reason why this is such an encouraging finding is that stellar
models played no role in the development of their color--$T_{\rm eff}$
relations.

Because its properties are so well determined (as noted at the beginning of 
this section), the Hyades provides the best available constraint on the color
transformations appropriate to super-metal-rich dwarf stars.  VC03 set up
color tables for [Fe/H] $=+0.3$ using the dependence of the various color
indices on [Fe/H] predicted by model atmospheres as a guide, and then refined
them so that interpolations between these tables and those for [Fe/H] $=0.0$
enabled a computed isochrone for the Hyades metallicity to match the cluster
CMD.  The result of this exercise is shown in Figure 4.  In this case, as well, 
there is independent support for the color--$T_{\rm eff}$ relations so obtained.
\citet{lcb98} had previously developed their own semi-empirical transformations,
completely independently of theoretical stellar models.  As shown by VC03, their
results also agree very well with the \citet{sf00} relation between $B-V$ and
$T_{\rm eff}$ for solar abundance stars --- and if their color table for [Fe/H]
$=+0.3$ is substituted for the one derived by VC03, and the isochrone
appropriate to the Hyades is transposed to the observed plane using the revised
color transformations, the result is the dashed curve in Fig.~4. The differences
between it and the solid curve are clearly not very significant.  The
color--$T_{\rm eff}$ relations for [Fe/H] $\ge 0.0$ by both VC03 and Lejeune et
al.~are appear to be quite robust.

\begin{figure}[!ht]
\plotone{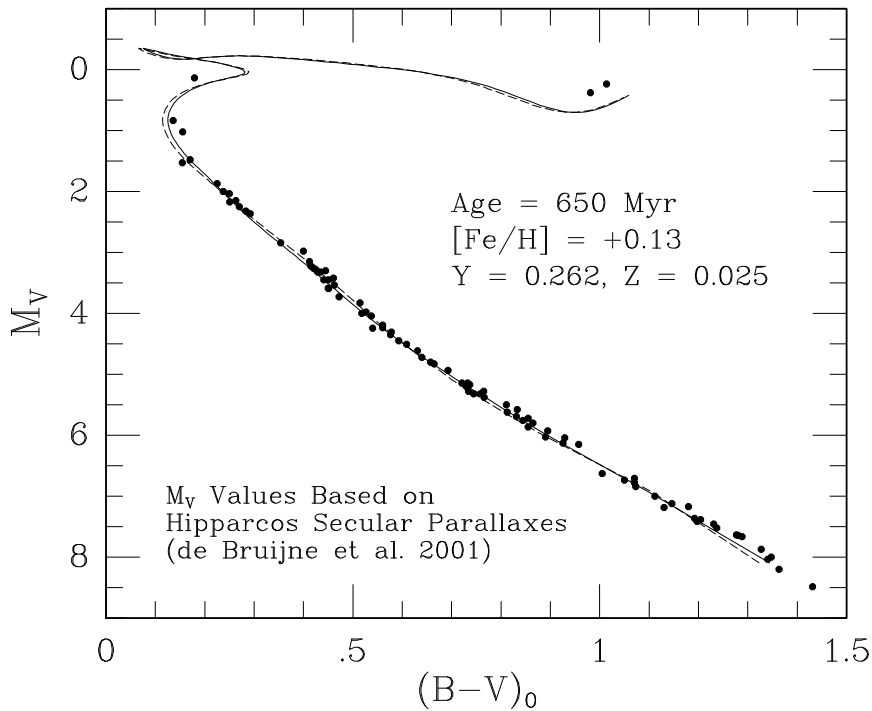}
\caption{Fit of an isochrone for the indicated parameters onto the de Bruijne
et al.~(2001) CMD for the ``high-fidelity" sample of single Hyades members.
The VC03 color transformations were used for the {\it solid curve}.
Interpolations between their table for [Fe/H] $=0.0$ and that by Lejeune et
al.~(1998) for [Fe/H] $=+0.3$ were used to calculate the colors along the
{\it dashed curve} (see the text).}
\end{figure}

\subsection{Comparison of Solar Evolutionary Tracks}
The calibration of the color transformations for stars having [Fe/H] $=0.0$ has
the distinct advantage that the temperature scale of stellar models can be
constrained using the Sun.  Ensuring that the models reproduce one point on
the H-R diagram does not, however, imply that they will agree elsewhere.
Differences in the assumed physics, boundary conditions, and the metal abundance
mixture, as well as the treatment (or not) of diffusive processes, among other
things, will all affect the values of $Y$ and $\alpha_{\rm MLT}$ that are
derived from a Standard Solar Model.  This (and any differences in the assumed
physics) will have implications for, e.g., the location of the giant branch of
a computed track relative to its MS location.  But just how large would one
expect these differences to be?

The answer to this question is given in Figure 5, which compares the
evolutionary tracks computed for the Sun by several researchers (as noted) using
their own codes.  (The goal here was not to compare the output of the different
evolutionary codes when as close to the same input physics is assumed, but
rather to illustrate the variations in the solar tracks that result from the net
effect of differences in the codes, adopted metallicities, and input
assumptions.)  Some relevant information about these tracks is provided in
Table 1, including, in particular, the age at the tip of the RGB.  Needless to
say, it is very comforting to find that the predicted ages agree so well. 

\begin{table}[!ht]
\caption{Solar Tracks Computed by Different Workers}
\smallskip
\begin{center}
{\small
\begin{tabular}{cccccc}
\tableline
\noalign{\smallskip}
Code & $Z$ & Diffusion? & $Y$ & $\alpha_{\rm MLT}$ & age \\
\noalign{\smallskip}
\tableline
\noalign{\smallskip}
 Cassisi & 0.01981 & yes & 0.2734 & 1.89 & 12.10 \\
 Girardi & 0.01900 & no  & 0.2730 & 1.68 & 12.25 \\
 VandenBerg & 0.01880 & no & 0.2768 & 1.90 & 12.18 \\
 Weiss   & 0.01998 & yes & 0.2754 & 1.74 & 12.07 \\
 Yi      & 0.01810 & yes & 0.2670 & 1.74 & 12.03 \\
\noalign{\smallskip}
\tableline
\end{tabular}
}
\end{center}
\end{table}

It is apparent in Fig.~5 that the largest differences occur near the tip of the
RGB, where the tracks span a range of $\approx 0.02$ in $\log T_{\rm eff}$ at
a fixed $M_{\rm bol}$.  Much of this can be attributed to the diversity in
the adopted values of $\alpha_{\rm MLT}$. For instance, the tracks computed by
Cassisi and by VandenBerg assume the highest value of the mixing-length
parameter and, consistent with expectations, their RGBs are hotter than those
computed by the other workers. (It is, however, a little surprising that these
two tracks agree so well given that one allows for diffusive processes, but
the other does not --- see Table 1.  There must be compensating effects due to
other factors; e.g., the assumed $Z$.)  The track by Girardi assumes the lowest
value of $\alpha_{\rm MLT}$, and it is the coolest track at the base of the RGB
(as it should be, all else being equal), though it ends up close the middle of
the five tracks near the RGB tip (which also indicates that other factors must
be having an impact). Still, the level of agreement is quite satisfactory
between the different computations.

\begin{figure}[!ht]
\plotone{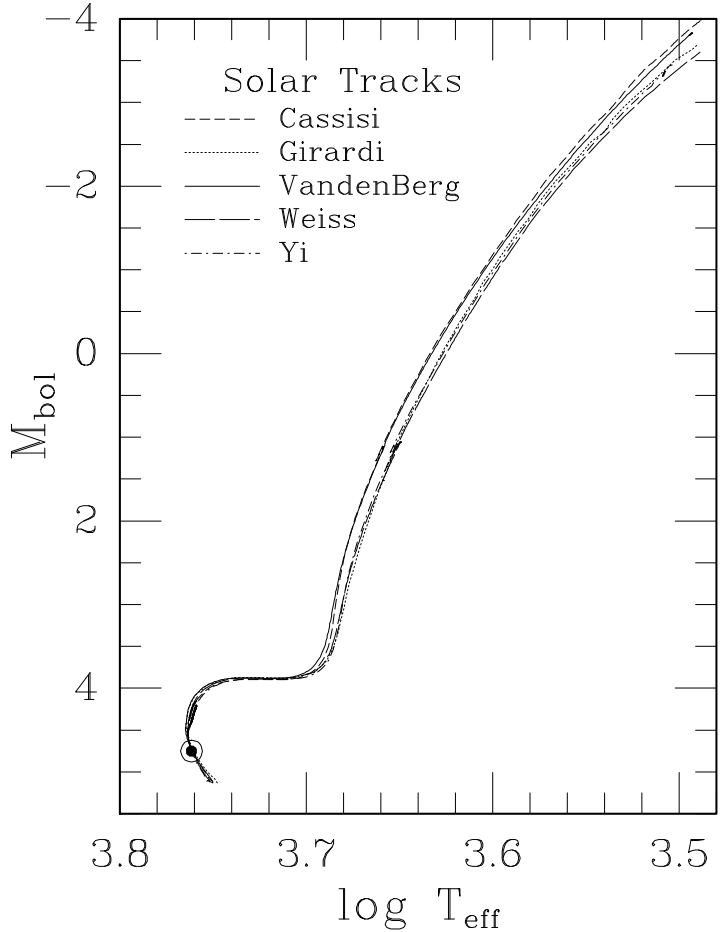}
\caption{Comparison of solar tracks computed the indicated researchers.  To aid
in their identification, the tracks, in the direction from left to right near
the RGB tip, were computed by S.~Cassisi, D.~VandenBerg, L.~Girardi, S.~Yi,
and A.~Weiss, respectively.  The Sun is represented by the solar symbol.}
\end{figure}

Indeed, all of the tracks probably comply with current constraints on the
temperatures of solar abundance giants to within their uncertainties.  They
should also be able to reproduce the CMDs of M$\,$67 and the Hyades quite
satisfactorily, simply by employing slightly different transformations than
those reported by VC03.  Indeed, it would be interesting to see how much of a
discrepancy between theory and observation would be found if the same
color--$T_{\rm eff}$ relations were used by all the different grids of
isochrones currently in use.  Of course, in the case of M$\,$67, in particular,
there is enough leeway in the chemical composition and reddening that
comparably good fits to the cluster photometry could well be obtained using
most of the available sets of models if, e.g., somewhat different metallicities
were assumed.

\section{The Calibration of Stellar Models for Metal-Deficient Stars}
Not having an extremely metal-poor counterpart to the Sun that can be similarly
used as a high-precision calibration point, or a low-metallicity star cluster
whose properties are known to exceedingly high accuracy, makes it very difficult
to assess the reliability of the temperatures and colors predicted by stellar
models at low $Z$.  As discussed below, the temperatures determined for field
Population II dwarfs having good distance estimates from {\it Hipparcos} can
sometimes vary by up to 100--150 K, depending on how they are derived.  This is
quite comparable to the uncertainties in $T_{\rm eff}$ that are found for field
and GC giants from infrared photometry.  Coupled with the $\sim 0.3$ dex
uncertainties in [Fe/H] that exist for many objects, it is relatively easy for
stellar models to satisfy many, if not most, observational constraints.  All
that may be needed to accomplish this is to choose a particular metallicity
scale --- e.g., that by \citet{zw84} in preference to the one by \citet{cg97},
or {\it vice versa}, in the case of Galactic GCs.  It is fair to say that
significant improvements to the prediction of the temperatures and colors of
metal-poor stars may not be possible until we can confidently measure their
chemical compositions to within at least $\pm 0.1$ dex.  We are still a long
ways from achieving that.

This section will bring to the fore some of the difficulties, apparent
inconsistencies, etc., that hinder further progress at the present time.  On
the one hand, it is very encouraging that model-atmosphere-based color--$T_{\rm
eff}$ relations appear to fine for solar metallicity stars warmer than $\sim
5000$--5500 K.  One would naively expect that they should do at least as well at
lower metal abundances given the concomitant decrease in the blanketing of
stellar atmospheres.  On the other hand, predicted temperatures undoubtedly
suffer from systematic errors that are not easily quantified.  Consequently, it
is not clear that stellar models which are transposed to the observational plane
using synthetic color--$T_{\rm eff}$ relations will accurately reproduce the
properties of real metal-poor stars.  (Even when there appears to be good
agreement between theory and observations, the possibility exists that errors
in the various factors that play a role have conspired to produce it.)

\begin{figure}[!ht]
\plotone{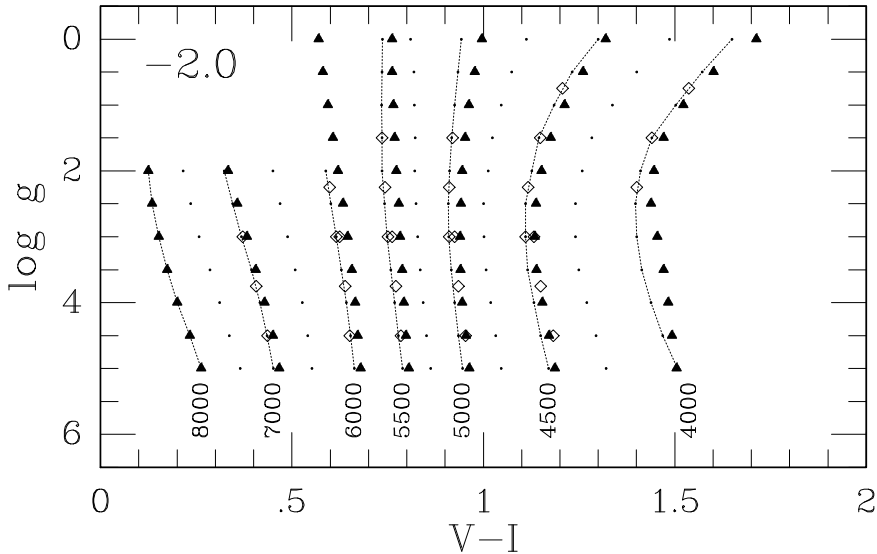}
\caption{Similar to Fig.~1, except that the color transformations are for
stars having [Fe/H] $=-2.0$.}
\end{figure} 

The obvious first step in testing the available color transformations that
apply to stars of low metallicity is to simply try them out.  As illustrated in
Figure 6, the $(V-I)$--$T_{\rm eff}$ relations derived from MARCS and Kurucz
model atmospheres for [Fe/H] $=-2.0$ agree quite well in a systematic sense,
though Kurucz colors are redder by $\sim 0.02$--0.03 mag over most of the ranges
in temperature and gravity that have been considered therein.  (The differences
in the predicted $B-V$ color indices from the two sources tend to be somewhat
larger, and to be stronger functions of $T_{\rm eff}$ and $\log g$; see Fig.~3
by VC03.)  It is difficult to say which of the two sets of transformations is
the most realistic one.  Comparisons of isochrones with observed CMDs may
indicate a clear preference, but that could well be more of a reflection on the
models than on the color transformations.  In effect, choosing to use a
particular set of color--$T_{\rm eff}$ relations from among several possible
alternatives is equivalent to {\it calibrating} the stellar models so that they
yield a particular interpretation of, e.g., the data for a few key stars or star
clusters.  (Only when the same model atmospheres are used to evaluate
the surface boundary conditions of the stellar interior models {\it and} to
accomplish their transformation to the various observational planes can it be
claimed that they represent a purely theoretical prediction.)

For this reason, it is important to choose the ``calibrating" objects very 
carefully.  Among the most metal-deficient GCs, M$\,$68 provides perhaps the
best benchmark because it has well-defined CMDs (\citealt{wal94}) based on
observations taken in three different bandpasses ($BVI$) during the same run
and subjected to the same reduction and standardization procedures.  Its
reddening is fairly low --- $E(B-V) = 0.06$ according to the Schlegel et
al.~(1998) dust maps --- and, at least until recently, its metallicity has not
been the subject of much controversy.  On the \citet{zw84} scale, it has [Fe/H]
$=-2.09$, while \citet{cg97} obtained [Fe/H] $=-1.99$ from high-resolution
spectroscopy of cluster giants.  Unfortunately, a metallicity near $-2.0$ has 
not been found by \citet{ki03} from their analysis of the equivalent widths
of Fe$\,$II lines: they obtained [Fe/H] $=-2.43$.  The consequent uncertainty
in the metal abundance of M$\,$68 obviously makes any comparison of its CMD
with stellar models rather insecure.

Still, it is worthwhile to proceed and to explore the implications of the
isochrones computed by \citet{bv01}, in particular, as an instructive example.
Throughout this study, the adopted distance modulus for any GC that is
considered will be based on a fit of a ZAHB locus calculated by \citet{vsr00}
to the lower bound of the distribution of cluster HB stars.  According to the
recent work by \citet{dc99} and by \citet{ccc05}, the luminosities of the RR
Lyrae stars in these systems, as derived from analyses of their pulsational
properties, are close to those predicted by the VandenBerg et al.~models.
Indeed, as will become apparent below, ZAHB-based distance estimates agree quite
well with those determined using other methods.

\subsection{M$\,$68}
If the evolutionary calculations for [$\alpha$/Fe] $=0.3$ and [Fe/H] $=-2.01$
are converted to the observational planes using the transformations from MARCS
and Kurucz model atmospheres, and overlayed onto the M$\,$68 CMDs assuming
$E(B-V) = 0.06$ and $(m-M)_V = 15.18$, the results shown in Figure 7 are 
obtained.  The solid curves, which employ the color--$T_{\rm eff}$ relations
from MARCS atmospheric models, clearly provide the best match to the cluster
CMD.  If Kurucz colors are used, the isochrones are generally too red (at least
for the adopted cluster parameters); and while most of the discrepancies on the
[$(V-I)_0,\,M_V$]--plane appear to be consistent with a small zero-point shift,
which would not be a serious concern, the fit of the isochrones to the cluster
fiducial on the [$(B-V)_0,\,M_V$]--diagram is much more problematic.  Granted,
the computed RGBs are still a bit too red when MARCS transformations are
adopted, but the differences are small enough that it might easily be a mistake
to attribute them to errors in the predicted temperatures and/or the
color--$T{\rm eff}$ relations.  The remaining discrepancies may instead be a
consequence of, e.g., the assumed cluster properties (reddening, distance,
chemical composition) not being quite right.

\begin{figure}[!ht]
\plotone{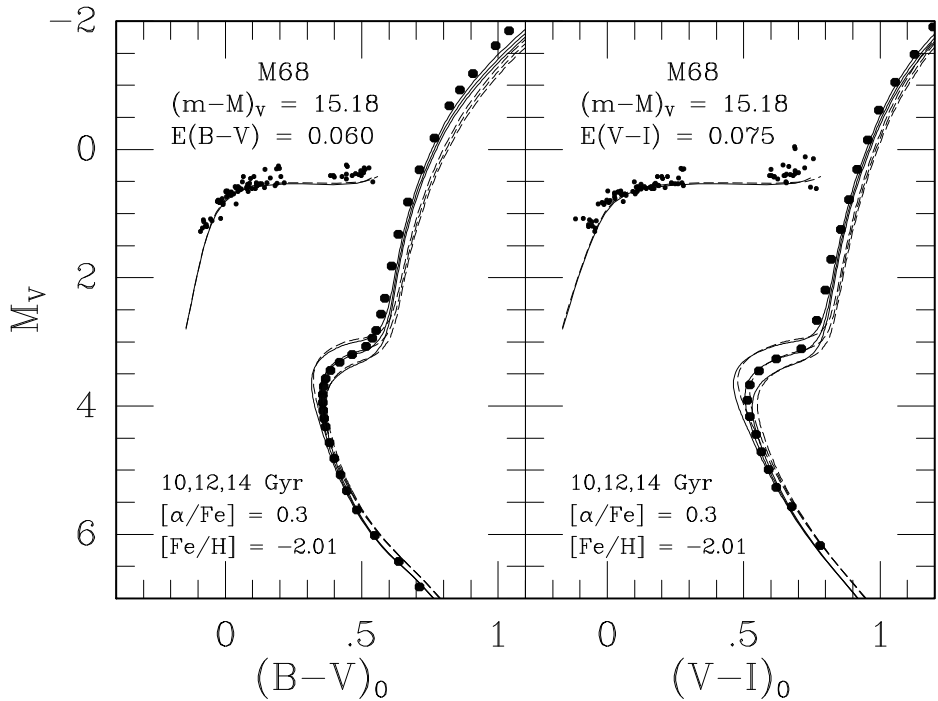}
\caption{Overlay of isochrones (Bergbusch \& VandenBerg 2001) and a fully
consistent ZAHB locus (VandenBerg et al.~2000) for the indicated parameters
onto the CMD of M$\,$68 by Walker (1994), as represented by the fiducial
sequences ({\it filled circles}) determined by DAV from the published
photometry.  The {\it solid} and {\it dashed curves} are obtained when the
color transformations from MARCS and Kurucz model atmospheres are used,
respectively.  Note that $E(V-I) = 1.25\,E(B-V)$ (Dean et al.~1978) has been
assumed.}
\end{figure}

It is noteworthy that very similar interpretations of the data are found in
both panels of Fig.~7 and that, if the distance is set using the HB, the lower
MS of M$\,$68 is well matched by the models.  [As pointed out by VC03,
comparable consistency is not obtained if the Lejeune et al.~(1998)
transformations are used.  Either there are some problems with the latter or
with the M$\,$68 observations.]  If, for instance, M$\,$68 is significantly more
metal poor than [Fe/H] $\approx -2.0$, then the HB stars should be somewhat
brighter if they conform to theoretical expectations, implying an increased
distance modulus, and the lower MS stars would lie above the ZAMS for the lower
metallicity.  In this case, redder color transformations would be needed.
To be sure, it is also possible that M$\,$68 actually has both a lower metal
abundance {\it and} a shorter distance than assumed, in which case, an even
better fit to the cluster CMD than that shown in Fig.~7 might be found using
the same MARCS color transformations.  (A lower metallicity isochrone for the
same or slightly higher age will have a bluer RGB as well as a reduced
difference in color between the turnoff and the RGB.)  These few comments serve
to illustrate some of the difficulties that are encountered when trying to
establish the color--$T{\rm eff}$ relations for very metal poor stars using
observational constraints that have appreciable uncertainties.

It goes without saying that any fine-tuning of the color transformations to
obtain a ``perfect" match to the CMD of M$\,$68 (or any other globular cluster)
cannot be justified.  What is encouraging is that, without applying any
corrections at all to the transformations from MARCS model atmospheres, it is
possible to get quite a respectable fit of isochrones to the entire CMD of
a very metal-deficient GC like M$\,$68, and to have a consistent interpretation
of both $BV$ and $VI$ data.  Indeed, VC03 found that similar good agreement
could be obtained for more metal-rich systems, provided that some adjustments
were made to the predicted colors for upper RGB and lower MS stars: the higher
the metallicity, the warmer the temperature at which these corrections seemed
to be necessary.  However, such a trend is not unexpected given that substantial
corrections to the color--$T_{\rm eff}$ relations are needed at rather warm
temperatures if models for [Fe/H] $\approx 0.0$ are to satisfy empirical
constraints (see \S 2).  Importantly, no adjustments were applied to the
theoretical color transformations at temperatures appropriate to turnoff stars
at any metallicity.  Indeed, the main effect of correcting the synthetic colors
for cooler stars is to ensure that the predicted RGB and MS slopes are in good
correspondence with those observed.   Models that fail to satisfy at least
these constraints will not be very useful for the interpretation of the
photometric data for complex stellar populations. 

\subsection{M$\,$92 and 47 Tucanae}
Let us now consider some comparisons of isochrones with the well-defined CMDs
for M$\,$92 and 47 Tuc that were obtained by \citet{sh88} by \citet{hhv87},
respectively.  In the lower panel of Figure 8, the fiducial sequences for these
two GCs have been superposed onto a number of \citet{bv01} isochrones for [Fe/H]
values ranging from $-2.31$ to $-0.40$ (assuming [$\alpha$/Fe] $=0.3$ in each
case) and ages that approximately reflect the GC age--metallicity relation
derived by \citet{van00}.  (To take into account the effects of diffusive
processes, which were not treated in the models that have been plotted, the
ages should be reduced by $\sim 10$\%; see VandenBerg et al.~2002.)  The adopted
reddenings are from the Schlegel et al.~(1988) dust maps, and the assumed
$(m-M)_V$ values follow from the fits of computed ZAHBs (VandenBerg et
al.~2000) to the cluster HB stars that are shown in the upper panel.  All of
the models have been transposed to the [$(B-V)_0,\,M_V$]--plane using the
{\it calibrated} color--$T_{\rm eff}$ relations given by VC03.

\begin{figure}[!ht]
\plotone{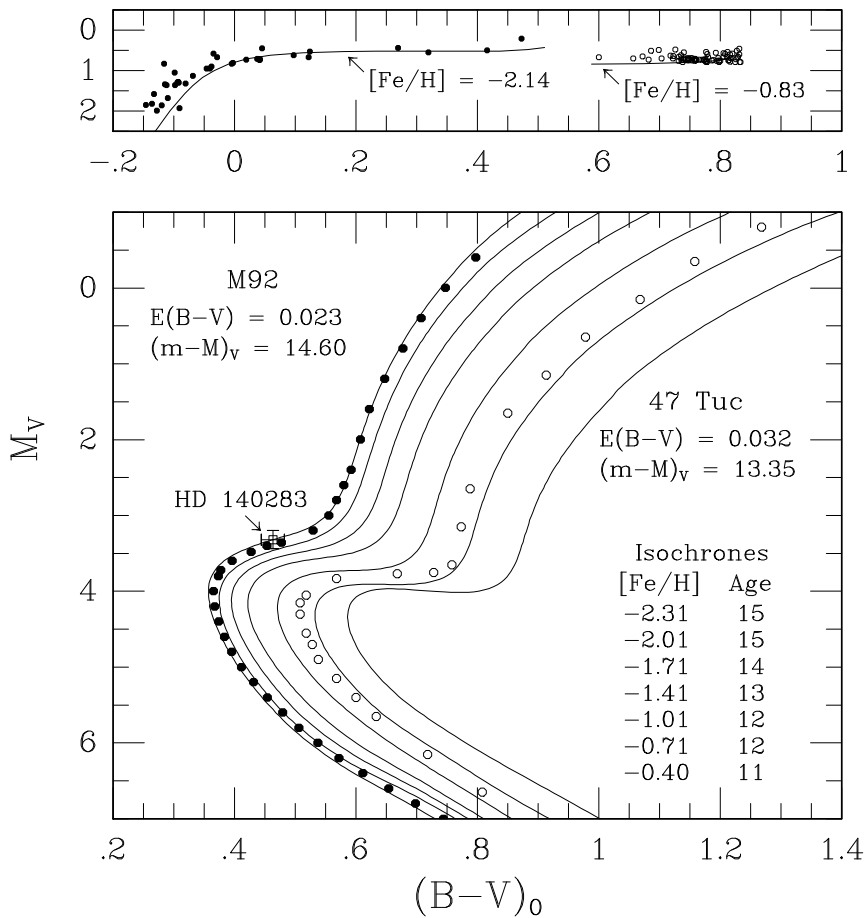}
\caption{The {\it lower panel} superimposes on a plot of isochrones, for the
indicated [Fe/H] values and ages, the fiducial sequences for M$\,$92 (Stetson
\& Harris 1988; {\it filled circles}) and 47 Tucanae (Hesser et al.~1987;
{\it open circles}) assuming the reddenings from the Schlegel et al.~(1998)
dust maps and distance moduli based on fits of computed ZAHBs (VandenBerg et
al.~2000) to the cluster counterparts, as shown in the {\it upper panel}.  The
location of the field Population II subgiant, HD$\,$140283, is given by the
{\it open square}.}
\end{figure}

Clearly, the predicted and observed RGB and MS slopes are in good agreement and,
moreover, there is excellent consistency between the ZAHBs and the lower MS
locations insofar as the assumed/implied [Fe/H] values are concerned.  (If, for
instance, {\it lower} metallicity ZAHBs were fitted to the cluster HB stars,
the consequent increase in the inferred distances would result in overlays of
the cluster MS fiducials onto isochrones of {\it higher} metal abundance ---
and {\it vice versa}.)  Thus, if the actual metallicities of M$\,$92 and 47 Tuc
are as assumed in Fig.~8, the fine agreement between theory and observations
which is evident therein would provide a strong argument in support of the
colors of the models that have been used in both an absolute and relative sense
(provided that, e.g., the photometric zero-points and reddenings are accurate).

Unfortunately, current estimates of the metallicity of M$\,$92 vary from [Fe/H]
$=-2.16$ (\citealt{cg97}) to $-2.38$ (\citealt{ki03}).  (An intermediate value,
$-2.24$, was obtained by \citealt{zw84}).  Even though there are good reasons
to favor the Kraft \& Ivans determination (e.g., being based on Fe II lines,
their results are largely unaffected by departures from LTE), it is not
impossible that M$\,$92 has a metallicity as high as [Fe/H] $=-2.14$, or even
higher.  Recent work on the local subgiant, HD$\,$140283, which is one of the 
best observed field Population II stars, has called into question metal
abundance determinations based on 1-D model atmospheres.  (The location of this
star has been plotted in Fig.~8 to support the ZAHB-based distance to M$\,$92.
This star should be slightly brighter than the cluster subgiants at the same
color, by approximately the amount shown, if it is the same age as, but slightly
more metal poor than, the globular cluster.)  Whereas nearly all spectroscopic
studies of HD$\,$140283 over the years have found [Fe/H] $\approx -2.4$ (see,
e.g., the [Fe/H] catalogue by \citealt{csf97}), \citet{sba05} have suggested
from model atmospheres which take 3-D effects into account that a reduction
of $\approx 100$ K in the $T_{\rm eff}$ usually adopted for this star together
with an assumed [Fe/H] $\approx -2.0$ result in much improved agreement between
the predicted and observed spectra.  It remains to be seen what the implications
of the more sophisticated model atmospheres will be for the metallicity of
M$\,$92, but an upward revision is conceivable.

The possible ramifications of 3-D model atmospheres notwithstanding, there is a
general consensus that 47 Tuc has an iron content very close to [Fe/H] $=-0.70$
(\citealt{zw84}; \citealt{cg97}; \citealt{ki03})  This poses a bit of a problem
for the models plotted in Fig.~8, which indicate that a consistent
interpretation of the HB and MS observations is obtained only if the cluster
metallicity is [Fe/H] $\approx -0.8$.  There are at least a couple of possible
ways of explaining this (admittedly small) discrepancy.  First, based on a
careful analysis of published photometry for 47 Tuc, using the secondary cluster
standards established by \citet{ste00}, \citet{psv02} have suggested that, in
the mean, the \citet{hhv87} observations are too blue by $\approx 0.02$ mag.
Indeed, an analogous plot to Fig.~8, but with the 47 Tuc fiducial shifted to  
the red by 0.02 mag, shows that the observed lower MS coincides very well with
a that of 12 Gyr, [Fe/H] $=-0.71$ isochrone if, as required by the fit of the
ZAHB for this metallicity to the cluster HB stars, $(m-M)_V = 13.30$ is
assumed.\footnote{Note that an apparent distance modulus of 13.30--13.35, as
inferred from HB models, is in excellent agreement with recent determinations
based on the cluster white dwarfs ($13.27\pm 0.14$; \citealt{zro01}) and on MS
fits to field subdwarfs ($13.33\pm 0.04\pm 0.1$; \citealt{gsa02}).}  (This
plot has not been included here because it is already quite obvious from Fig.~8
that such a redward adjustment of the 47 Tuc CMD would result in a close match
to the isochrone for [Fe/H] $=-0.71$.)

Errors in the adopted reddening or the model colors could also explain why the
deduced [Fe/H] value from Fig.~8 is at variance with spectroscopic metallicity
estimates.  For instance, \citet{gbc03} have recently obtained $E(B-V) = 0.024
\pm 0.004$ for 47 Tuc, which is $\approx 0.01$ mag smaller than the value
assumed here.  This supplies approximately one-half of the shift needed to
reconcile the cluster CMD with models for [Fe/H] $\approx -0.7$.  While it is
quite possible that the remaining difference can be traced to the
color--$T_{\rm eff}$ relations that have been used, it is just as likely (if
not more so) that the isochrones are too red because the effective temperatures
predicted by the models for metal-poor stars are too low.  As discussed by
\citet{vsr00}, stellar models that are computed using MARCS model atmospheres
as boundary conditions will be $\sim 50$--100 K warmer (depending on the metal
abundance and evolutionary state) than those which assume a scaled-solar
$T$--$\tau$ relation to describe the atmospheric structure.

\begin{figure}[!ht]
\plotone{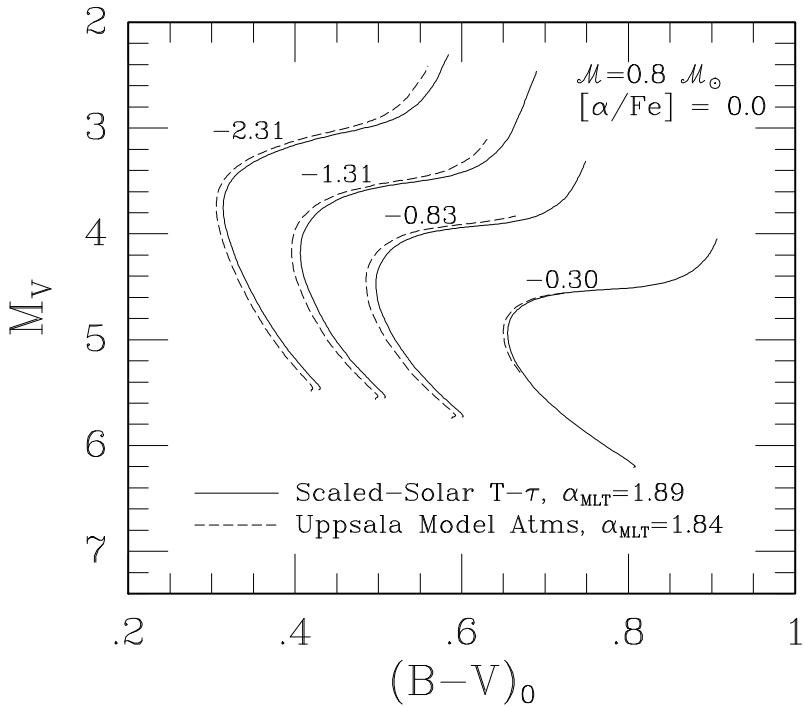}
\caption{Tracks for $0.8 {{\cal M}_\odot}$ having [$\alpha$/Fe]$=0.0$ and the
[Fe/H] values specified adjacent to the various loci, in which the boundary
conditions are derived either from model atmospheres ({\it dashed curves}) or by
integrating the hydrostatic equation in conjunction with the Krishna Swamy
(1966) scaled-solar $T$--$\tau$ relation ({\it solid curves}). Note that
slightly different values of the mixing-length parameter, as indicated, are
implied by a Standard Solar Model, and that the dashed curves are less
complete than the solid curves because model atmospheres were available only
for $T_{\rm eff} \ge 5500$ K.}
\end{figure} 
 
Their results are presented on the [$(B-V)_0,\,M_V$]--plane in Figure 9.  This
shows that evolutionary tracks for $0.8 {{\cal M}_\odot}$ stars having [Fe/H]  
$=-2.31$, $-1.31$, $-0.83$, and $-0.30$ are 0.005--0.02 mag bluer if the
surface boundary conditions of the individual stellar models are derived from
model atmospheres rather than from integrations of the hydrostatic equation
in conjunction with the \citet{ks66} $T$--$\tau$ relation.  (This prediction
should be regarded as ``preliminary" as it is based on a small number of model
atmospheres that were made available $\sim 6$ years ago by B.~Gustafsson and
B.~Edvardsson.  A collaborative project with the Uppsala group is presently
in progress to more fully analyze the impact of using their latest model
atmospheres to describe the outermost layers of stellar models.)  The degree
to which isochrones, as opposed to evolutionary tracks, are affected remains
to be seen, but it seems probable that predicted turnoff color versus age
relations will be bluer by one- or two-hundredths of a magnitude (at a fixed
age) if the effects of proper model atmospheres are taken into account.  

While such offsets seem desirable for our understanding of 47 Tuc (at the
present time, anyway), they would appear to pose a problem for the most
metal-deficient systems.  Fig.~9 suggests that all of the isochrones in Fig.~8
more metal poor than $\sim -0.7$ should be shifted to the blue by at least
0.01 mag.  However, any blueward correction of the isochrones appropriate to
M$\,$92 (and M$\,$68, for that matter) would have the consequence that a
consistent explanation of its CMD (i.e., of both its HB and MS stars) would
be possible only if a somewhat higher metallicity were assumed (closer to
[Fe/H] $=-2.0$ than to $-2.14$), which would conflict with current
spectroscopic estimates.  To be sure, with so many sources of uncertainty, it
is easily possible that errors in one or more quantities are compensating for
errors in others.  Being able to predict the colors of stars of any
metallicity to within $\sim 0.02$ mag could well be the best that we can do
(and will do for the foreseeable future), and once that point is reached,
further debate may not be very meaningful.  Realistically, one cannot say
with any certainty that there is, or is not, a problem with the comparison 
between theory and observations presented in Fig.~8.

\subsection{M$\,$3 and M$\,$5}
\citet{sbh99} fiducials for M$\,$3 and M$\,$5 are compared with a set of
12 Gyr isochrones spanning the range in metallicity, $-2.14\le$ [Fe/H] $\le
-1.14$, in Figure 10, which is very similar to the lower panel in Fig.~8
except that the [$(V-I)_0,\,M_V$]--plane is considered.  The fiducial
sequence for M$\,$68 from Fig.~7 has been replotted here to reiterate that it
can be reproduced quite well by models for [Fe/H] $\approx -2.0$ --- i.e.,
for the metal abundance derived by \citet{cg97}.  The assumed reddenings are
from the Schlegel et al.~(1998) dust maps and, although not shown, the adopted
distance modulus in each case is such that a theoretical ZAHB, for that
metallicity which is implied by the overlay of the predicted and observed lower
main sequences, provides a good fit to the lower bound of the distribution of
cluster HB stars.  Thus, the models provide consistent fits to the MS and
HB stars in M$\,$3 and M$\,$5 only if their metallicities are [Fe/H] $\approx
-1.6$ and $-1.4$, respectively.

\begin{figure}[!ht]
\plotone{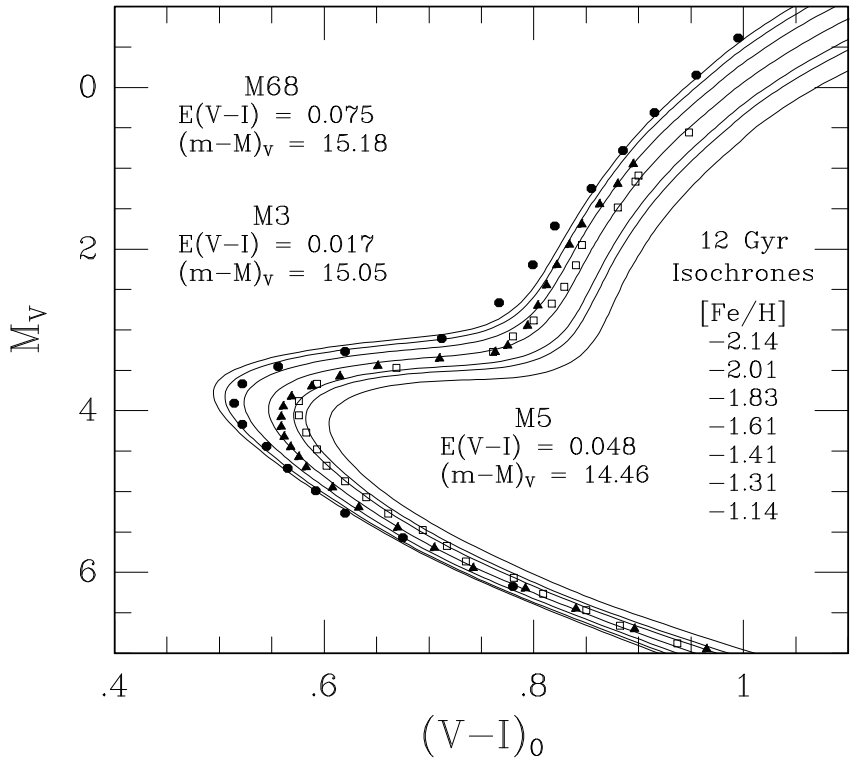}
\caption{Similar to the lower panel in Fig.~8, except that 12 Gyr isochrones
for the indicated [Fe/H] values are compared with the fiducial sequences of
M$\,$3 ({\it filled triangles}) and M$\,$5 ({\it open squares}) from Stetson
et al.~(1999).  The same fiducial of M$\,$68 that appeared in Fig.~7 has been
replotted here as {\it filled circles}.}
\end{figure}

This is an intriguing result because these estimates are not in good agreement
with the [Fe/H] values derived by \citet{cg97} using high-resolution
spectroscopy (even though there is no such conflict in the case of the most
metal-deficient clusters, like M$\,$68).  According to Carretta \& Gratton,
the iron abundance of M$\,$3 is [Fe/H] $=1.34$, while that of M$\,$5 is
$-1.11$.  However, it is well known that, although the \citet{zw84} and
\citet{cg97} [Fe/H] scales are very similar below [Fe/H] $\approx -2.0$ and
above $\approx -1.0$, they tend to differ by $\sim 0.3$ dex at intermediate
metallicities.  In fact, the [Fe/H] values of M$\,$3 and M$\,$5 as inferred
from the models (see Fig.~10) agree very well with the Zinn \& West
determinations; namely, $-1.66$ and $-1.40$, respectively.  Interestingly,
metal abundances derived from Fe II lines are close to the mean of the
Carretta--Gratton and Zinn--West results: \citet{ki03} have obtained [Fe/H]
$=-1.50$ for M$\,$3 and $-1.26$ for M$\,$5.  (Consistency with the findings
of Kraft \& Ivans would be obtained if the isochrones were adjusted to the
blue by as little as 0.01--0.015 mag.)

It is clearly very hard to assess the accuracy of the temperatures and colors
of stellar models from comparisons of isochrones with observed CMDs until, at
the very least, metallicity determinations are placed on a much firmer
footing.  Even then, uncertainties in the basic cluster properties (distance,
reddening, and photometric zero-points) limit what can be done.  Perhaps the
main value of Fig.~10 is to show how precisely everything must be known in
order to obtain the correct interpretation (whatever that may be) of any given
CMD: note that the separation in the lower MS segments of the isochrones at
$M_V\ge 5.5$ is only $\sim 0.075$ mag in $V-I$ for a 1 dex change in [Fe/H].

\subsection{Globular Cluster Standard Field Photometry (Stetson 2000)}
When \citet{van00} determined the ages of a sample of 26 GCs, ZAHB-based
distance moduli and reddenings from the Schlegel et al.~(1998) dust maps were
adopted (as in this study).  On the basis of these assumptions, the observed
CMDs were converted to [$(B-V)_0,\,M_V$]- or [$(V-I)_0,\,M_V$]--diagrams.  The
age of each cluster was then obtained by first adjusting the isochrones for
the assumed metallicity in color until the predicted MS overlaid that
observed, and then noting which isochrone best reproduced the location of the
turnoff and the subgiant branch.  The advantage of this approach is that
errors in the model temperatures or colors, or problems with the photometric
zero-points or reddening, do not affect the derived age.  As noted in the
plots provided by VandenBerg, the model colors were sometimes adjusted to the
blue and sometimes to the red, by typically $\sim 0.02$ mag.  Moreover, the
corrections needed to fit $BV$ and $VI$ photometry for the same cluster were
not always in the same direction.  Only for a few clusters, like M$\,$68
(\citealt{wal94}), was it possible to obtain consistent interpretations of
the data on two different color planes.  (As already noted, Walker's $BVI$
observations were taken in the same observing run, and subjected to the same
reduction and calibration procedures.)

\citet{rsp99}, among others, have emphasized the importance of having a
homogeneous photometric database (in their case, to obtain self-consistent
relative GC ages).  Indeed, the lack of homogeneity could well be the main
reason why there was no obvious pattern to the corrections that \citet{van00}
applied to the isochrones in his investigation.  Needless to say, the
possibility of zero-point and/or systematic errors in cluster photometry is
a real concern for any attempts that are made (e.g., that by VC03) to use  
such data to calibrate synthetic colors.  Fortunately, P.~B.~Stetson (see
\citealt{ste00}) is engaged in a monumental, on-going effort to homogenize
a large fraction of the $BVRI$ CCD photometry that has been acquired by many
researchers over the years.  In particular, he is in the process of
establishing many ``standard fields" in GCs that can be used for calibration
purposes.  Given their availability, an obvious question to ask is: ``How well
are \citet{bv01} isochrones, using VC03 color transformations, able to
reproduce the CMDs defined by his secondary standards in, say, 47 Tuc, M$\,$5,
and M$\,$92?".\footnote{The data for these clusters were downloaded from the
Canadian Astronomy Data Center web site (http://cadcwww.hia.nrc.ca/standards)
on April 25, 2005.}

The answer to this question, as revealed in Figure 11, is ``not as well as
one would have hoped".  As far as 47 Tuc is concerned, the predicted MS (on
both color planes) is about 0.01 mag too blue, suggesting that models for a
slightly higher [Fe/H] value than $-0.83$ would be more appropriate.
(Isochrones for [Fe/H] $=-0.71$, which is the next highest metallicity in our
grid, provided a less satisfactory fit to the observations than that shown.)
Curiously, the RGB segments of the isochrones are 0.01--0.02 mag too blue on
the [$(B-V)_0,\,M_V$]--plane, but too red by about the same amount on the
[$(V-I)_0,\,M_V$]--diagram.  It is quite possible that these discrepancies 
are due to minor problems with the color transformations, particularly those
for the $B-V$ color index.   

\begin{figure}[!ht]
\plotone{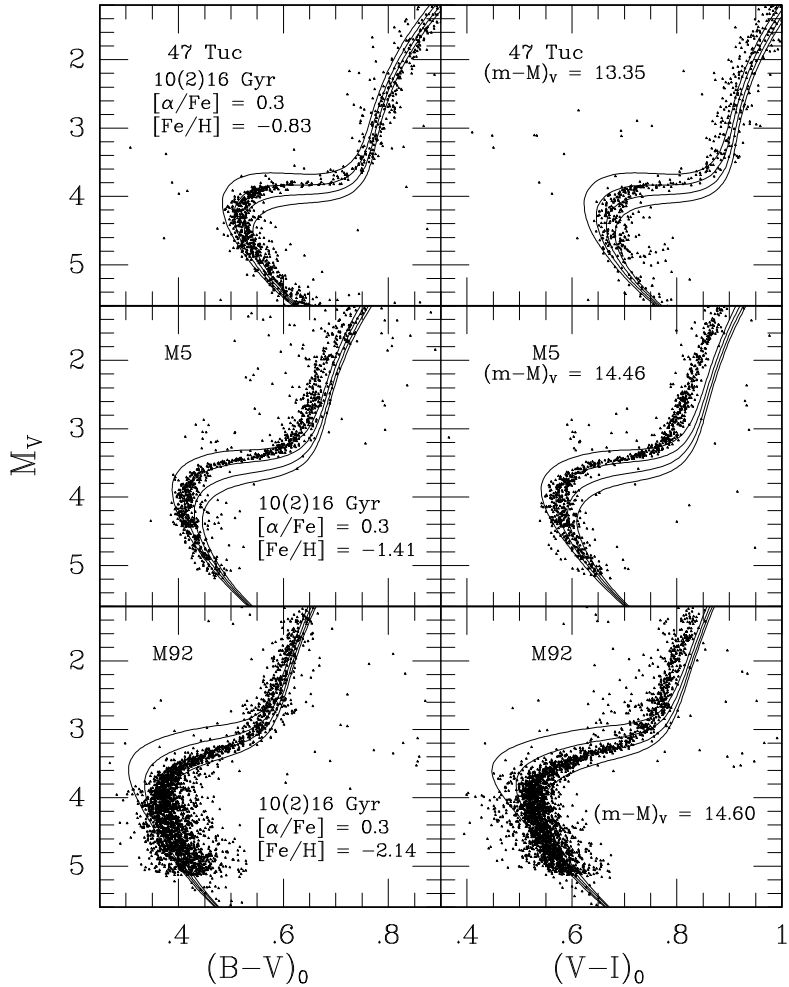}
\caption{Comparisons of isochrones for the indicated ages and chemical
compositions with the ``standard cluster field" CMDs (Stetson 2000) for
47 Tuc, M$\,$5, and M$\,$92.  For the three clusters, the same reddenings and
distance moduli are assumed as in Figs.~8 and 10.}
\end{figure}

The offsets between the computed and observed RGBs are much bigger ($\approx
0.04$ mag) in the case of M$\,$5.  They are larger than any found in the
extensive study of GC CMDs by \citet{van00}, though it must be acknowledged
that most of his fits of isochrones to photometric data were discrepant in the
same sense.  It seems unlikely that the color--$T_{\rm eff}$ relations
are at the root of this difficulty because the predictions from MARCS model 
atmospheres, which were adopted without appreciable modifications by VC03
for [Fe/H] values $\le -1.0$, tend to be bluer than other transformations
currently in use (e.g., see Fig.~6).  Either the model temperatures along
the giant branch are too cool, which is possible (though they appear to be
consistent with empirical constraints, as shown in the next section), or our
understanding of M$\,$5 is somehow lacking (assuming, of course, that the
CMDs of M$\,$5 presented in Fig.~11 are trustworthy).  Isochrones for a lower
metallicity, a higher oxygen abundance, or an increased age should, for
instance, provide a better match to the observed (small) difference in color
between the TO and the RGB.

It is, perhaps, especially disheartening that isochrones which had previously
provided a superb match to the CMD of M$\,$92 by Stetson \& Harris (1988; see
Fig.~8) do not fare nearly as well when confronted by the standard
field data.  As shown in the lower left-hand panel of Fig.~11, the observed
MS lies slightly to the red of the isochrones while the cluster giants are,
in the mean, $\sim 0.02$ mag bluer than the models.  Of particular concern is
the fact that, in the [$(V-I)_0,\,M_V$]--diagram (see the lower right-hand
panel), the deviation between the predicted and observed MS of M$\,$92 is in
the opposite sense, and the discrepancies along the giant branch are
appreciably larger (though in the same direction as in the left-hand panel).
Thus, the models do not provide consistent interpretations of the $BV$ and
$VI$ observations.  Does this mean that Walker's (1994) photometry for M$\,$68
suffers from systematic errors (given that, as shown in Fig.~7, no such
inconsistency is found for this cluster, which has a very similar metal
abundance as M$\,$92)?

There are clearly some mysteries here that need to be understood.  Perhaps the
main conclusion to be drawn from this brief examination of the standard field
data is that, when using the CMDs for GCs to test the predicted colors of
stellar models, one must be wary of the very real possibility that the
observations suffer from significant zero-point and/or systematic errors.
 
\subsection{Giant-branch Temperatures}
A valuable test of the predicted temperatures of giants is provided by
empirical $(V-K)$--$T_{\rm eff}$ relations.  Due to the extensive work by
\citet{fpc83}, fiducial sequences on the $(\log T_{\rm eff},\,M_bol)$--plane
have been derived for many star clusters from $V-K$ photometry.  Those 
reported by \citet{fpc81} for M$\,$92, M$\,$3, 47 Tuc, and M$\,$67 are plotted
in Figure 12, together with the RGB segments of the isochrones that appeared
in Fig.~8.  (The distance moduli adopted by Frogel et al have been corrected
by $\approx 0.2$ mag, or less, to be consistent with the ZAHB-based distances
that have been adopted throughout this study.  Because the giant branches 
rise so steeply, such adjustments have no more than minor consequences for
the comparisons with theoretical models.)  It is immediately apparent that
the predicted and ``observed" RGB slopes are nearly the same, which suggests
that the value of the mixing-length parameter, $\alpha_{\rm MLT}$, does not
vary by much, if at all, during the first ascent of the giant branch.
Furthermore, the empirically derived temperatures for M$\,$92, M$\,$3, and
47 Tuc are well reproduced by the models for [Fe/H] $\approx -2.3$, $-1.55$,
and $-0.7$, respectively, though the size of the $1\sigma$ error bar is large
enough to accommodate larger or smaller [Fe/H] values by $\sim 0.15$--0.3 dex.
There is clearly no conflict with current spectroscopic determinations, and
hence no indication that $\alpha_{\rm MLT}$ varies with metallicity.  Indeed,
much tighter constraints on the temperatures of giants are needed if they are
to provide more exacting tests of stellar models.

\begin{figure}[!ht]
\plotone{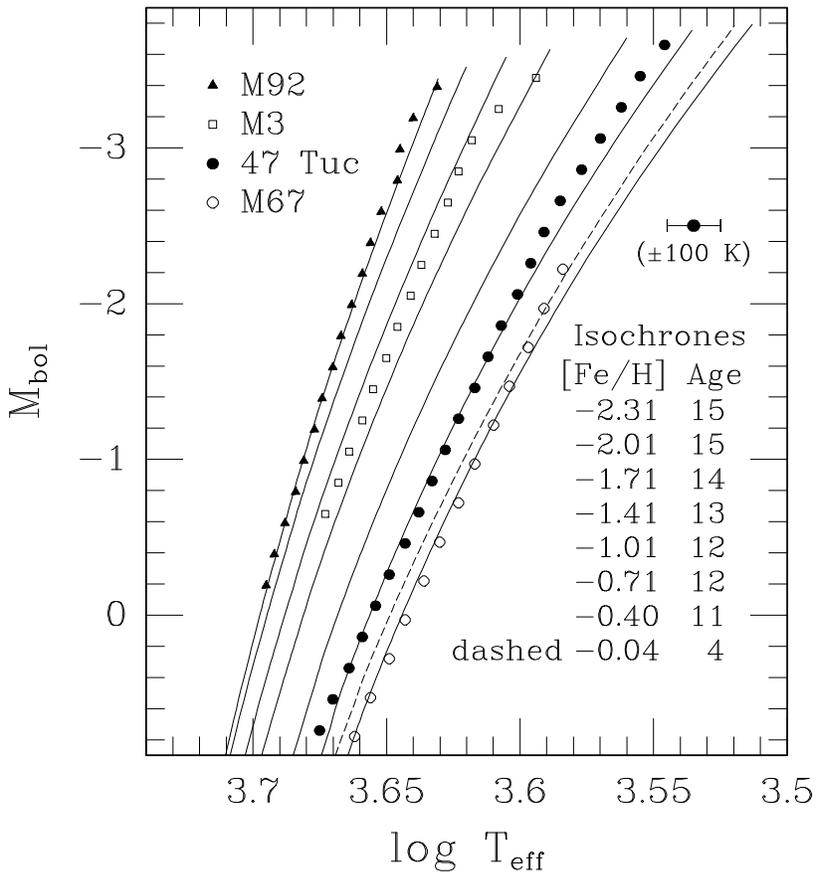}
\caption{Comparison of the upper RGB segments of isochrones for the indicated
[Fe/H] values and ages with the fiducial sequences for M$\,$92, M$\,$3, 47
Tuc, and M$\,$67 that were derived by Frogel et al.~(1981) using infrared
photometry.  The {\it solid curves} assume [$\alpha$/Fe] $=0.3$, whereas the
{\it dashed curve} assumes scaled-solar abundances.  The distance moduli
adopted by Frogel et al.~have been revised slightly (see the text).}
\end{figure}

As mentioned in \S 2 (also see VC03), the predicted temperatures along the
giant-branch segment of the 4 Gyr isochrone used to fit the M$\,$67 CMD (see
Fig.~2) agree well with those determined from $V-K$ photometry and the latest
available $(V-K)$--$T_{\rm eff}$ relations (Bessell et al.~1998; van Belle et
al.~1999).  The upper part of this isochrone has also been plotted in Fig.~12
(the {\it dashed curve}) in order to determine how much it differs from the
temperatures derived for M$\,$67 giants by Frogel et al.~(1981; the {\it open
circles}) nearly 25 years ago.  The agreement between the two is obviously
quite good, which indicates that the transformation between $V-K$ and
$T_{\rm eff}$ has withstood the test of time rather well.

\subsection{Population II subdwarfs}
Although our comparisons of isochrones with the CMDs of M$\,$3 and M$\,$5
(see Figs.~10 and 11) indicate a preference for the \citet{zw84} [Fe/H] scale
over that by \citet{cg97} in the intermediate metal-poor regime, this is not
a consequence of the color--$T{\rm eff}$ relations that have been used.
Indeed, the VC03 transformations are completely consistent with those used by
R.~G.~Gratton and colleagues to determine the temperatures of the Population
II subdwarf standards with [Fe/H] values between $-1.0$ and $-2.0$ that have
often been used to derive GC distances via the MS fitting technique.  As shown
in Figure 13, the $B-V$ colors that are determined from interpolations in the
VC03 color tables for the values of $T_{\rm eff}$ and [Fe/H] derived by
\citet{gcc96} and \citet{cgc99} for some of the best observed subdwarfs (also
see \citealt{bv01}), using $\log g$ values from stellar models, are in
exceedingly good agreement with the observed colors.  (In the case of dwarf
stars, colors are not very sensitive to the assumed gravity; consequently,
very similar results would obtained for any $\log g$ value between 4.3 and
4.7.)
 
\begin{figure}[!ht]
\plotone{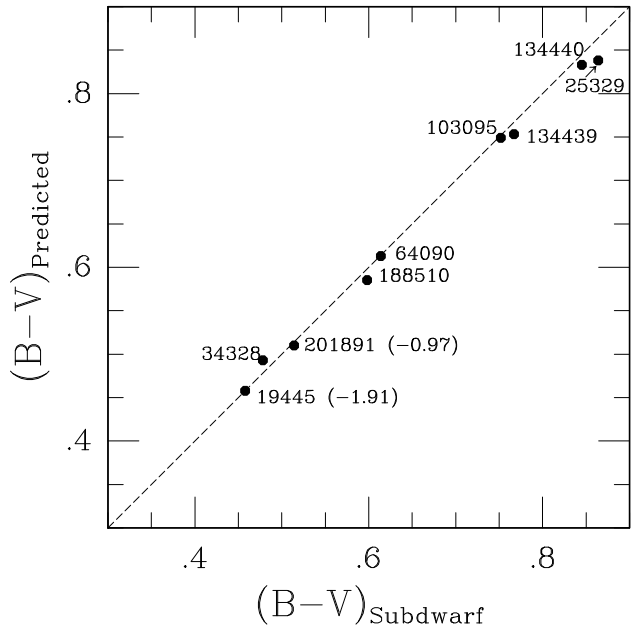}
\caption{Comparison of the predicted and observed colors of several Population
II subdwarfs that are identified by their HD numbers.  The former are derived
by interpolations in the VC03 color tables using the temperatures and
metallicities (which range between [Fe/H] $=-0.97$ and $-1.91$, as indicated)
determined for the subdwarfs by Gratton et al.~(1996) and Clementini et
al.~(1999), and using values of $\log g$ from stellar models.}
\end{figure}

The agreement is remarkable considering that (i) the assumption of higher or
lower temperatures by as little as 50 K would affect the predicted colors by
$\sim 0.02$ mag for the coolest stars, and (ii) the temperatures derived for 
the same star from color--$T_{\rm eff}$ relations (e.g., Gratton et al.~1996),
the infrared-flux method (e.g., \citealt{aam96}), the fitting of Balmer line
profiles (e.g., \citealt{afg94}), and fits to the {\it uv}--flux distributions
(e.g., \citealt{al00}) sometimes differ by up to $\sim 150$ K.  (However, for
many of the stars considered in Fig.~13, the temperatures obtained via the
different techniques agree quite well --- see Table 5 by Clem et al.~2004.)
To the extent that the assumed properties of the subdwarfs have been 
accurately estimated, Fig.~13 provides additional support for the color
transformations championed by VC03.

\subsection{Comparison of $0.8 {{\cal M}_\odot}$ Tracks for [Fe/H] $=-2.3$}
Figure 14 presents a comparison of the computed tracks from completely
independent stellar evolution codes for a very low value of [Fe/H].  As for
the calculations shown in Fig.~5 for solar abundances, there are a number
of differences in the assumed chemistry and the adopted physics, some of
which are noted in Table 2.  (The assumed values of $\alpha_{\rm MLT}$ are
the same as in Table 1.)  The main intent of such a comparison is to
determine how much of an effect the different assumptions and physics have
on, in particular, the predicted temperatures.

\begin{figure}[!ht]
\plotone{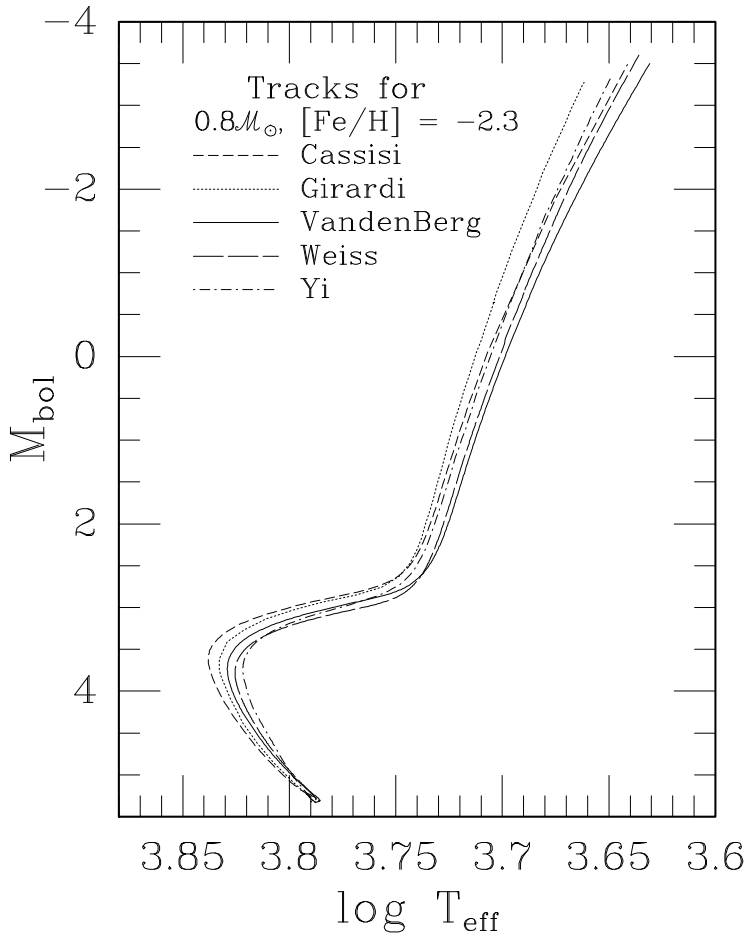}
\caption{Similar to Fig.~5, except that tracks for $0.8 {{\cal M}_\odot}$
stars having [Fe/H] $=-2.3$ are compared.}
\end{figure}
 
\begin{table}
\caption{$0.8 {{\cal M}_\odot}$ Tracks for [Fe/H] $=-2.3$ Computed by
 Different Workers}
\smallskip
\begin{center}
{\small
\begin{tabular}{cccccc}
\tableline
\noalign{\smallskip}
Code & $Z$ & Diffusion? & $Y$ & $\alpha$-element & age \\
     &     &            &     &   enhanced?  &   \\
\noalign{\smallskip}
\tableline
\noalign{\smallskip}
 Cassisi & 0.0001 & no & 0.2450 & no & 12.58 \\
 Girardi & 0.0001 & no  & 0.2300 & no & 13.53 \\
 VandenBerg & 0.0001 & no & 0.2352 & no & 13.22 \\
 Weiss   & 0.0001 & no & 0.2351 & no & 13.74 \\
 Yi      & 0.0001 & yes & 0.2300 & yes & 13.48 \\
\noalign{\smallskip}
\tableline
\end{tabular}
}
\end{center}
\end{table}

In fact, the differences are larger than one might have anticipated. 
Curiously, the Girardi and VandenBerg tracks coincide almost exactly between
the ZAMS and the lower RGB, but they diverge thereafter: Girardi's track has
the warmest giant branch of all those plotted in Fig.~14, while VandenBerg's
is the coolest.  [Without a detailed examination of the respective codes that
were used, it is not possible to explain the cause(s) of these differences.]
On the one hand, models that predict warmer RGBs should do a better job of
matching GC CMDs if the VC03 color--$T_{\rm eff}$ relations are used to
transform the models to the observed plane.  On the other hand, the empirical
$T_{\rm eff}$ scale for giants is supportive of the VandenBerg predictions
(as shown in \S 3.5) --- though the uncertainties are probably large enough
to encompass all of the theoretical models that have been plotted.  It is not
surprising that the track computed by Yi has the coolest TO given that his
calculations were the only ones to take diffusion into account.  On the whole,
there seems to be reasonable consistency between the different computations;
e.g., the age is approximately inversely correlated with the assumed He
abundance and the subgiant luminosity, as expected.

\section{Conclusions}
For simple stellar populations like those found in open and globular star
clusters, it really does not matter very much if there are problems with the
colors of the isochrones that are compared with the observations.  Distance
determinations are considered to be far more reliable, for instance, if they
are based on so-called ``standard candles" (lower MS stars having accurate
distances and metallicities, RR Lyraes, white dwarfs) than on fits to stellar
models.  Indeed, it may be possible to find an isochrone that appears to
provide a very good match to all of the principal sequences in an observed CMD, 
but that isochrone would probably not be the correct explanation of the data.
There are too many factors that affect the shapes of isochrones, including the
overall metallicity, the detailed heavy element mixture (notably of the CNO and
$\alpha$ elements), the treatment of convection, the low-temperature opacities,
the color--$T{\rm eff}$ relations, physics usually not taken into account
(e.g., turbulence at the base of the envelope convection zone, rotation),
and more.  Each of these variables has its own uncertainty and, collectively,
they will certainly permit substantial variations in e.g., the predicted slope
of the subgiant branch, or the difference in color between the TO and the lower
RGB.  It should not be too surprising, or too disturbing, to find that stellar
models do not provide as satisfactory a fit to color-magnitude data as one
would like.  Fortunately, TO luminosities are subject to far fewer (and better
understood) uncertainties, and they can be used to obtain well-constrained ages
for those systems having good distance estimates and metal abundances that have
been determined spectroscopically.  
 
The situation is quite different in the case of complex stellar populations;
i.e., systems containing stars that span wide ranges in age and chemical 
composition.  In order to unravel their star formation and chemical evolution
histories, both the luminosity and color distributions in the observed CMDs
must be explained.

\acknowledgements      
Sincere thanks go to Santi Cassisi, Leo Girardi, Achim Weiss, and Sukyoung Yi
for providing the evolutionary tracks, computed using their respective codes,
that have been plotted in Figures 5 and 14.  I am particularly grateful to
David Valls-Gabaud for giving me the opportunity to present this paper.  The
support of a Discovery Grant from the Natural Sciences and Engineering Research
Council of Canada is also gratefully acknowledged.


\end{document}